\documentclass{aa}  

\usepackage{graphicx}
\usepackage{txfonts}
\begin{document} 
\newcommand{\ETACOLD}{\eta_{rc}}
\newcommand{\ETAHOT}{\eta_{rh}}
\newcommand{\TCOLD}{T_{rc}}
\newcommand{\THOT}{T_{rh}}

\setcounter{table}{1}

   \title{Lines and continuum sky emission in the near infrared: 
observational constraints from deep high spectral resolution spectra 
with GIANO-TNG\thanks{Tables 1, 2, and 4 are
only available at the CDS via anonymous ftp to
cdsarc.u-strasbg.fr (130.79.128.5) or via
http://cdsarc.u-strasbg.fr/viz-bin/qcat?J/A+A/vol/page}}

\titlerunning{The sky emission in the near infrared}
\authorrunning{E. Oliva, L. Origlia, S. Scuderi et al.}

   \author{E. Oliva \inst{1}   
          \and L. Origlia \inst{2}  
          \and S. Scuderi\inst{3} 
	  \and S. Benatti \inst{4} 
	  \and I. Carleo \inst{4} 
          \and E. Lapenna\inst{5} 
          \and A. Mucciarelli\inst{5} 
	  \and C. Baffa \inst{1}
          \and V. Biliotti \inst{1}
          \and L. Carbonaro \inst{1}
          \and G. Falcini  \inst{1}
          \and E. Giani\inst{1}  
          \and M. Iuzzolino\inst{1}  
          \and F. Massi\inst{1}
          \and N. Sanna\inst{1}  
          \and M. Sozzi\inst{1}
          \and A. Tozzi\inst{1}
          \and A. Ghedina\inst{6}  
          \and F. Ghinassi\inst{6}  
          \and M. Lodi\inst{6}
          \and A. Harutyunyan\inst{6}
          \and M. Pedani\inst{6}
          }

   \institute{
              INAF - Osservatorio Astrofisico di Arcetri,
              Largo E. Fermi 5, I-50125 Firenze, Italy\\ 
              \email{oliva@arcetri.inaf.it}
         \and
	     INAF - Osservatorio Astronomico di Bologna,
             via Ranzani 1, I-40127 Bologna, Italy
	 \and
	     INAF - Osservatorio Astrofisico di Catania,
	      via S. Sofia 78, I-95123 Catania, Italy
	 \and
	      INAF - Osservatorio Astronomico di Padova,
	       Vicolo Osservatorio 5, I-35122 Padova, Italy
	\and
             Universit\`{a} di Bologna, Dipartimento di Fisica e Astronomia, 
             Viale Berti Pichat 6/2, I-40127 Bologna, Italy
	 \and
             INAF - Fundaci\'{o}n Galileo Galilei,
             Rambla Jos\'{e} Ana Fern\'{a}ndez P\'{e}rez 7,
             E-38712 Bre\~{n}a Baja, TF, Spain
             }

   \date{Received .... ; accepted ...}

 
  \abstract
{}
{
Determining the intensity of lines and continuum airglow emission in the 
H-band is important for the design of faint-object infrared
spectrographs. Existing spectra at low/medium resolution cannot
disentangle the true sky-continuum from instrumental effects (e.g.
diffuse light in the wings of strong lines).  
We aim to obtain, for the first time, a high resolution
infrared spectrum deep enough to set significant constraints on the 
continuum emission between the lines in the H-band.
}
{
During the second commissioning run of the GIANO high-resolution 
infrared spectrograph at the La Palma Observatory,  
we pointed the instrument directly to the sky and obtained 
a deep spectrum that extends from 0.97 to 2.4 $\mu$m.
}
{
The spectrum shows about 1500 emission lines, a factor of two more than in
previous works. Of these, 80\% are identified as 
OH transitions; half of these are from highly excited molecules (hot-OH 
component) that are not included in the
OH airglow emission models normally used for astronomical applications.
 The other lines are attributable to O$_2$ or unidentified.
Several of the faint lines are in spectral regions that 
were previously believed to be free of line emission.
The continuum in the H-band is marginally detected
at a level of about 300 photons/m$^2$/s/arcsec$^2$/$\mu$m,
equivalent to 20.1 AB-mag/arcsec$^2$.
The observed spectrum and the 
list of observed sky-lines are published in electronic
format.  
}
{
Our measurements indicate that the sky continuum in the H-band 
could be even darker than previously believed. However, the myriad
of airglow emission lines severely limits the spectral ranges where very
low background can be effectively achieved with low/medium resolution 
spectrographs.
We identify a few spectral bands that could still remain quite dark
at the resolving power foreseen for VLT-MOONS (R$\simeq$6,600).\\
}

 \keywords{Line: identification -- Instrumentation: spectrograph -- Infrared: general -- 
   Techniques: spectroscopic }

   \maketitle

%

%

\section{Introduction}

The sky emission spectrum  at infrared wavelengths and up to 1.8 $\mu$m
(Y, J, H bands) is dominated by lines (airglow) emitted by OH and
O$_2$ molecules; see e.g. Sharma \cite{Sharma}. 
These lines are intrinsically very
narrow and, when observed at a high enough spectral resolution, they occupy only a 
small fraction of the spectrum. Therefore, by filtering the lines out, one
could in principle decrease the sky background by orders of magnitudes,
down to the level set by the sky continuum emission in between the lines.
This apparently simple idea, often reported as "OH sky-suppression",
has fostered a long and active field of research; see e.g. Oliva \& Origlia 
\cite{Oliva92}, Maihara et al. \cite{Maihara93}, Herbst \cite{Herbst94},
Content \cite{Content96}, Ennico et al. \cite{Ennico98},
Cuby et al. \cite{cuby00},
Rousselot et al. \cite{Rousselot00},
Iwamuro et al. \cite{Iwamuro01}, Bland-Hawthorn et al. \cite{blandhawthorn04},
Iwamuro et al. \cite{Iwamuro06},
Ellis et al. \cite{ellis12}, Trinh et al. \cite{Trinh13}.
However, in spite of the intense work devoted to measuring and modelling
the properties of the sky spectrum, it is still
not clear what is the real level of the sky continuum in between the
airglow lines in the H-band (1.5-1.8 $\mu$m).

A detailed study of the infrared sky continuum emission 
was recently reported by Sullivan \& Simcoe \cite{Sullivan}. Using
spectra at a resolving power R=6,000 they were able
to correct the spectra for all instrumental effects and derive accurate
measurements of the sky continuum at wavelengths shorter than 1.3 $\mu$m 
(Y, J bands). However, they could not obtain precise results
in the H-band (1.5-1.8 $\mu$m) because the sky continuum is well below the
light diffused in the instrumental wings of the airglow lines. 
This problem was already noted in earlier works.
In particular, Bland-Hawthorn et al. 
\cite{blandhawthorn04} claimed that the continuum level
 between the OH lines could be as low as the zodiacal light level and
much lower than that measurable with classical (i.e. not properly 
OH suppressed) spectrographs.
This claim was later retracted by Ellis et al. \cite{ellis12} after 
measuring the interline continuum with an optimised
OH-suppression device based on a Bragg fibre grating. 
Trinh et al. \cite{Trinh13} subsequently attempted to model
the interline continuum based on spectral models and measurements that did 
not reach the depth and completeness of the data presented in this paper.

The net - and somewhat surprising - result is that so far nobody has been 
able to improve the earliest measurements
of Maihara et al. \cite{Maihara93} who reported a continuum emission
of 590 photons/m$^2$/s/arcsec$^2$/$\mu$m measured at 1.665 $\mu$m
(equivalent to 19.4 AB-mag/arcsec$^2$)
using a spectrometer with resolving power R=17,000.
equipped with one of the first-generation 256$^2$ HgCdTe infrared detectors

A proper understanding of the line and continuum emission from the sky
is of fundamental importance when designing new infrared spectrographs
optimised for observations of very faint targets. A representative case
is that of MOONS, the multi-objects optical and near infrared spectrometer
for the VLT, see Cirasuolo et al.~(\cite{cirasuolo11}, \cite{cirasuolo14}).
This instrument includes an arm covering the H-band at a resolving power
R$\simeq$6,600. The requirements on instrumental background and stray light 
strongly depend on the sky continuum one assumes, see Li Causi et al.
\cite{licausi} for details.

In a previous work (Oliva et al. \cite{paper1}) 
we presented, for the first time, observations of the infrared sky spectrum
at high spectral resolution and covering a very wide spectral range.
The spectrum revealed 750 emission lines, many of these never reported
before. However, the data were not deep enough to provide significant
constraints on the continuum emission in between the lines.

Here we present and discuss new measurements taken with GIANO during the
second commissioning run at Telescopio Nazionale Galileo (TNG). In 
Sect.~\ref{observations} we briefly describe the instrument, the measurements,
and the data reduction. In Sects.~\ref{results} and \ref{discussion}
we present and discuss the results.

   \begin{figure*}
   \includegraphics[width=\hsize]{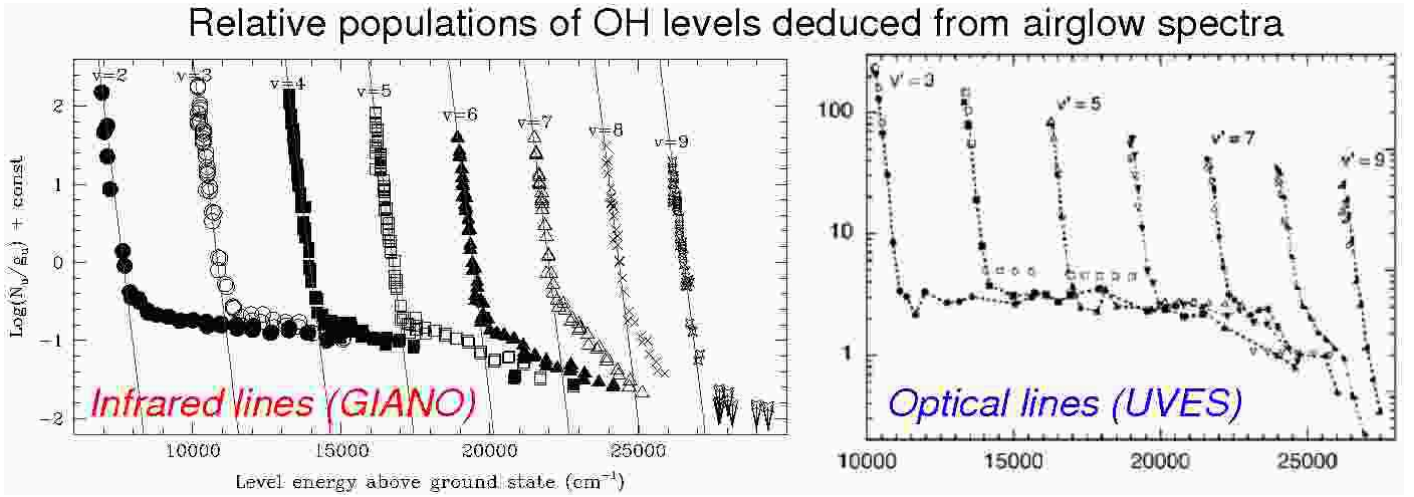}
      \caption{
        Derived column densities of the OH levels plotted against the
        energy of the levels above the ground-state of the molecule.
        The left panel shows the values derived from the infrared
        lines discussed here while the right hand panel -- reproduced
        under permission from
        Fig.~16 of Cosby \& Slanger \cite{Cosby};
        \copyright\ Canadian Science Publishing or its licensors
        -- summarizes the results based
        on optical (UVES) spectra.
        The steep straight lines in the left panel
        show the distribution predicted by standard
        models with rotational levels thermalised at 200~K.
        The quasi-flat tails reveal the hot-OH component, see
        Sect.~\ref{OH_hot} for details.
        }
        \label{fig_OH_levels_single}
   \end{figure*}
   \begin{figure*}
   \includegraphics[width=\hsize]{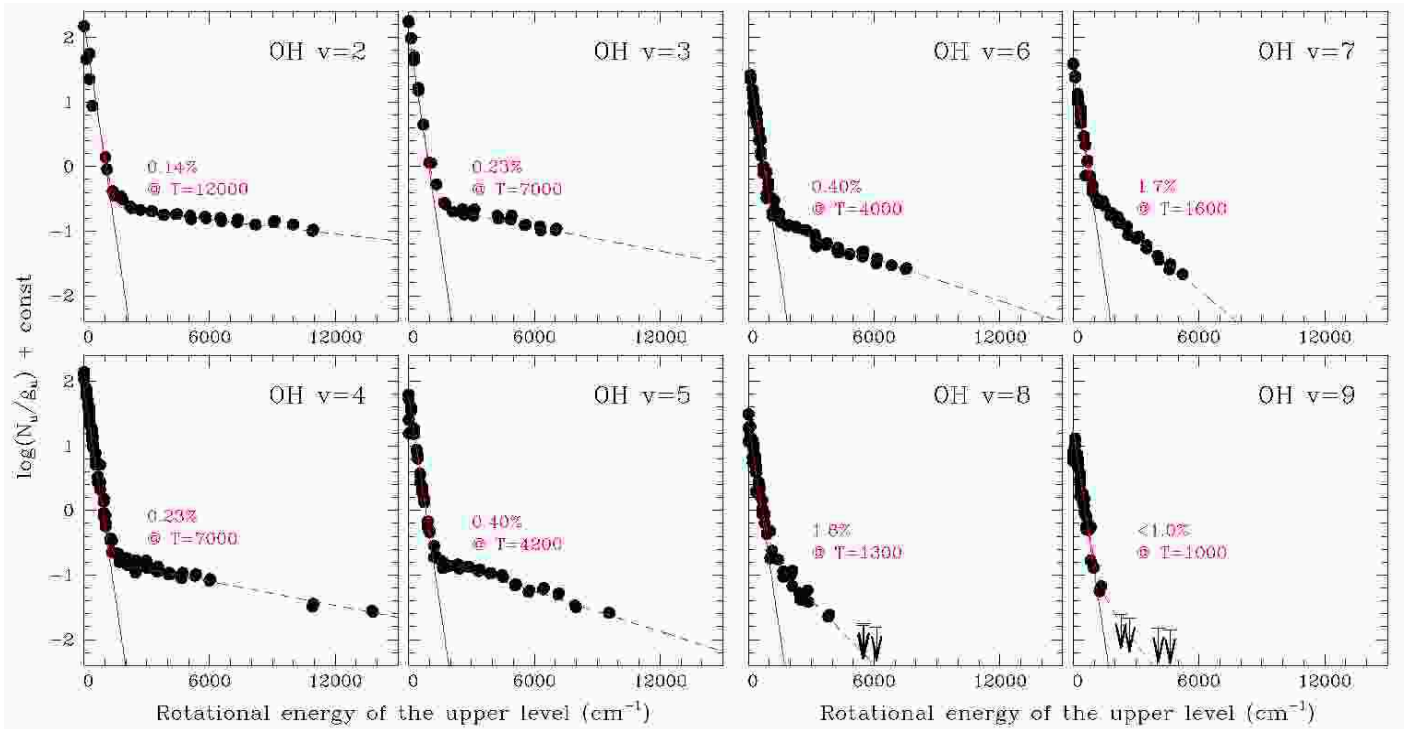}
      \caption{
        Same as Fig.~\ref{fig_OH_levels_single} but with separate
        panels for each vibrational level. The straight solid lines
        represent the cold-OH component while the dashed curves show the
        distribution obtained adding a fraction of hot-OH molecules.
        The numerical fraction and rotational temperature of the
        hot-OH molecules is reported within each panel.
        See text, Sect~\ref{OH_hot} for details.  }
         \label{fig_OH_levels_details}
   \end{figure*}

\section{Observations and spectral analysis}
\label{observations}
GIANO is a cross-dispersed cryogenic spectrometer that simultaneously 
covers the spectral range from 0.97~$\mu$m to 2.4~$\mu$m with a
maximum resolving power of R$\simeq$50,000 for a 2-pixels slit.  
The main disperser is a commercial R2 echelle grating
with 23.2 lines/mm that works on
a $\oslash$100 mm collimated beam. Cross dispersion is performed
by prisms (one made of fused silica and two made of ZnSe) that work in 
double pass. The prisms
cross-disperse the light both before and after it is dispersed by
the echelle gratings;
this setup produces a curvature of the images of the spectral orders.
The detector is a HgCdTe Hawaii-II-PACE with 2048$^2$ pixels. Its control
system is extremely stable with a remarkably low read-out noise 
(see Oliva et al. ~\cite{oliva_spie2012_2}).
More technical details on the instrument can be found in 
Oliva et al.\ ~\cite{oliva_spie2012_1} and references therein.

GIANO was designed and built for direct light feeding from the TNG 
3.5~m telescope.
Unfortunately, the focal station originally reserved 
was not available when GIANO was commissioned. Therefore we were
forced to position the spectrograph on the rotating building and develop
a complex light-feed system using a pair of IR-transmitting ZBLAN fibres 
with two separate opto-mechanical interfaces. 
The first interface is used to feed the telescope light into
the fibres; it also includes the guiding camera and the calibration
unit. The second interface re-images the light from the fibres onto
the cryogenic slit; see Tozzi et al.\ \cite{tozzi14} for more details. 

The overall performances of GIANO are negatively affected by 
the complexity of the interfaces and by problems intrinsic to the
fibres: the efficiency has been lowered by almost a factor of 3 and the
spectra are affected by modal-noise, especially at longer wavelengths.
Consequently, the observations of the sky taken during normal
operations are not appropriate to reveal the faintest airglow lines and
the continuum emission in between. To overcome this problem we took 
advantage of the early part of a test night (September 3, 2014) to secure
a direct spectrum of the sky with GIANO. We moved the spectrograph to the 
main entrance of the TNG and arranged a simple pre-slit system (a lens
doublet and two flat mirrors) to feed the cryogenic slit with the light from
a sky-area in the ESE quadrant at a zenith distance of about 25 degrees.
The half moon was in the SSW quadrant at a zenith distance of 50 degrees.

We integrated the sky for two hours using the 3-pixels slit that is normally 
used in combination with the fibre interface. This yields a resolving
power R$\simeq$32,000.
For calibration we took several long series 
of darks interspaced with flat frames. This strategy was chosen because 
the sky spectra showed some residuals (persistency) of a flat frame taken 
many hours before. We therefore re-created different pseudo-darks with
different levels of persistency and, during the reduction, we selected the 
combination of pseudo-darks that best reproduced the persistency pattern.
The criterion for selection relied on the assumption
that the sky continuum emission is zero in the spectral regions were the 
atmosphere is opaque, i.e. in the 1.37-1.40~$\mu$m range
(orders 54-56 of the GIANO echellogram). In other words,
we took advantage of the fact that the flat and its residual persistency
have similar intensities over the full spectral range, while the spectrum of
any light source above the troposphere (i.e. astronomical
targets and the sky airglow emission) is absorbed by the water bands.

The acquisitions were performed using the standard setup
of the controller, i.e. on chip
integrations of 5 minutes with multiple non-destructive read-outs every
10 seconds. All the read-outs were separately stored. The "ramped-frames"
were constructed later-on using the algorithm described in 
Oliva et al.~\cite{oliva_spie2012_2} that, besides applying the standard 
Fowler sampling, it also minimises the effects of 
reset-anomaly and cosmic rays.

The 1D spectra were extracted by summing 20 pixels along the slit.
Wavelength calibration was performed using U-Ne lamp frames taken
after the series of darks. The spectrometer is stable to $<$0.1 pixels
(i.e. $\Delta\lambda/\lambda\!<\!10^{-6}$ r.m.s.).
The wavelengths of the uranium lines were taken from Redman
et al.~\cite{redman2011}, while for neon we used the table available
on the NIST website\footnote{physics.nist.gov/PhysRefData/ASD/lines\_form.html}.
The resulting wavelength accuracy was about 0.07~\AA\ r.m.s. for lines in 
the H-band.

The flat exposures were used to determine and correct the variation of
instrumental efficiency within each order. An approximate flux calibration was
performed by assuming that the relative efficiencies of the orders are
the same as when observing standard stars through the fibre-interface 
and the TNG telescope. This is a very reasonable assumption within the
relatively narrow wavelength range covered by the H-band. However, it may
cause systematic errors (up to 0.3 dex) in the relative fluxes of lines 
with very different wavelengths.
Absolute flux calibration was roughly estimated by imposing that the flux
of the OH [4-0]Q1(1.5) line at 1.5833 $\mu$m is 270 photons/m$^2$/s/arcsec$^2$;
i.e. the typical value measured during normal observing nights.

\section{The sky lines and continuum emission}    
\label{results}
A total of about 1500 airglow lines were detected in the spectrum.
Compared to Paper~1, we have doubled the number of emission features measured.
In the following we separately discuss the OH lines, the other emission
features and the continuum emission in the Y, J, and H bands.
\subsection{OH lines and the hot-OH component}
\label{OH_hot}

Table~1 (available only in electronic format) lists
the lines identified as OH transitions.
For each $\Lambda$-doublet we give the wavelengths (in vacuum)
and the total observed flux of the doublet, normalised to the brightest
transition. 
For the fluxes we assumed that the two components of each
doublet have equal
intensities, i.e. that the '$e$' and '$f$' sub-levels are in thermal
equilibrium; this is appropriate for the density and temperature of the
mesosphere.
The listed wavelengths are derived from the  
newest OH molecular constants by Bernath \& Colin \cite{Bernath}.
These include highly excited rotational states and
allowed us to identify OH lines from rotational levels as high
as J=22.5, thus adding important constraints on the hot component 
of OH emission. 
This component was already reported by Cosby \& Slanger
\cite{Cosby} and in Paper~1. It is not included in any of the
models of OH airglow emission normally used for astronomical applications. 
These assume that the OH molecules have
a very high vibrational temperature ($T_{vib}\simeq 9000$~K) and a much lower
rotational temperature ($T_{rot}\simeq 200$~K).
In other words they assume that the gas density is high enough
to make collisional transitions between rotational states much faster
than radiative de-excitations. This brings the rotational temperature
to values similar to the kinetic temperature of the gas.
The net result is that all the lines from levels with rotational quantum 
number J$>$8.5 are normally predicted to be extremely faint and totally
negligible. 
The number of lines that are missed by standard models can be directly
visualised in Fig.~\ref{fig_OH_levels_single} that plots the column densities
of the upper levels of the measured lines as a function 
of the excitation energy of the levels. The steep lines show the distribution
expected for a single gas component with rotational levels thermalised
at T=200~K. 
The points in the quasi-flat tails represent emission lines
from hot molecules that are not thermalised. 
According to Cosby \& Slanger \cite{Cosby}, this hot 
component  is related to low density clouds at higher altitudes.
Here the gas density is lower than the critical density of the rotational
levels and, therefore, the population of the levels remain similar to that
set at the moment the OH molecule is formed.

In order to provide a practical tool to predict the intensities of all OH
lines we have fitted the observed level distribution with a mixture
of two components. The first is the standard model (cold-OH), while the 
second (hot-OH)
has a rotational temperature that is empirically determined from the observed
values.  Each vibrational state must be separately fit to obtain a good
matching.  
This simple model works as follows: let $N_u$ (cm$^{-2}$) 
be the column density of a given state ($v,J,F$) of the
OH molecule. This quantity is related to the excitation
temperatures by the standard Boltzmann equations, i.e.
$$ {N_u\over g_u N_{OH} } = e^{-E_v/k T_v} \ \ 
 \left[ \ETACOLD {e^{-E_{J,F}/k\TCOLD}\over U(T_v,\TCOLD)} 
 + \ETAHOT {e^{-E_{J,F}/k\THOT}\over U(T_v,\THOT)} \right] \eqno{(1)} $$
where $g_u$ is statistical weight of the level, $N_{OH}$
is the total column density of OH molecules, $E_v$ is 
the vibrational
energy of the level, $T_v$ is the vibrational temperature, $E_{J,F}$ is 
the rotational energy of the level, $\TCOLD$ is the rotational temperature 
of the cold component, $\ETACOLD$ is the fraction of cold molecules,
$\THOT$ is the rotational temperature of the hot
component, $\ETAHOT$ is the fraction of hot molecules
 and $U(T_v,T_r)$ is the partition function.
The photon-flux of a given transition arising from the same level is given by 
$$ I_{ul} = N_u \cdot A_{ul} \eqno{(2)} $$
where $A_{ul}$ (s$^{-1}$) is the transition probability.
The points in Figs.~\ref{fig_OH_levels_single}, \ref{fig_OH_levels_details}
are computed from Eq.~(2) using the observed line intensities together with 
the molecular parameters
of Bernath \& Colin \cite{Bernath} and the transition probabilities 
of van der Loo \& Groenenboom \cite{vanderloo}.
The steep straight lines in the left panel of Fig. ~\ref{fig_OH_levels_single} 
plot the function defined in
Eq.~(1) for $T_v$=9000~K, $\TCOLD$=200~K and $\ETACOLD$=1 (i.e. only cold-OH).
The same function is displayed in Fig. ~\ref{fig_OH_levels_details}
where the dashed curves show the results obtained adding a hot-OH component
with parameters ($\ETAHOT$,$\THOT$) adjusted for each vibrational level; the
values of the parameters are indicated in each panel.

The hot-OH component is most prominent in the lowest vibrational 
state (v=2) and becomes progressively weaker and cooler going to higher
vibrational states; it virtually disappears at v=9.

\subsection{ O$_2$ and unidentified lines }
\label{sect_O2}
The lines that cannot be associated with OH transitions are listed in
Table~2 (this table is available only in electronic format).
For the identification of the O$_2$ lines we used 
the HITRAN database (Rothman et al. \cite{hitran}). 
Most of the identified transitions were already reported in Paper~1.
A comparison between the two spectra shows that the intensity ratio 
between O$_2$ and OH lines has varied by almost a factor of 2 between
the two epochs. This is not surprising: the Oxygen lines are known
to vary by large factors even on timescales of hours. In our case
the variation can be used to select those features that follow the
time-behaviour of the O$_2$ lines. These lines are identified as  
``O2?'' (i.e. probably O$_2$) in Table~2.

\begin{figure*}
   \includegraphics[width=\hsize]{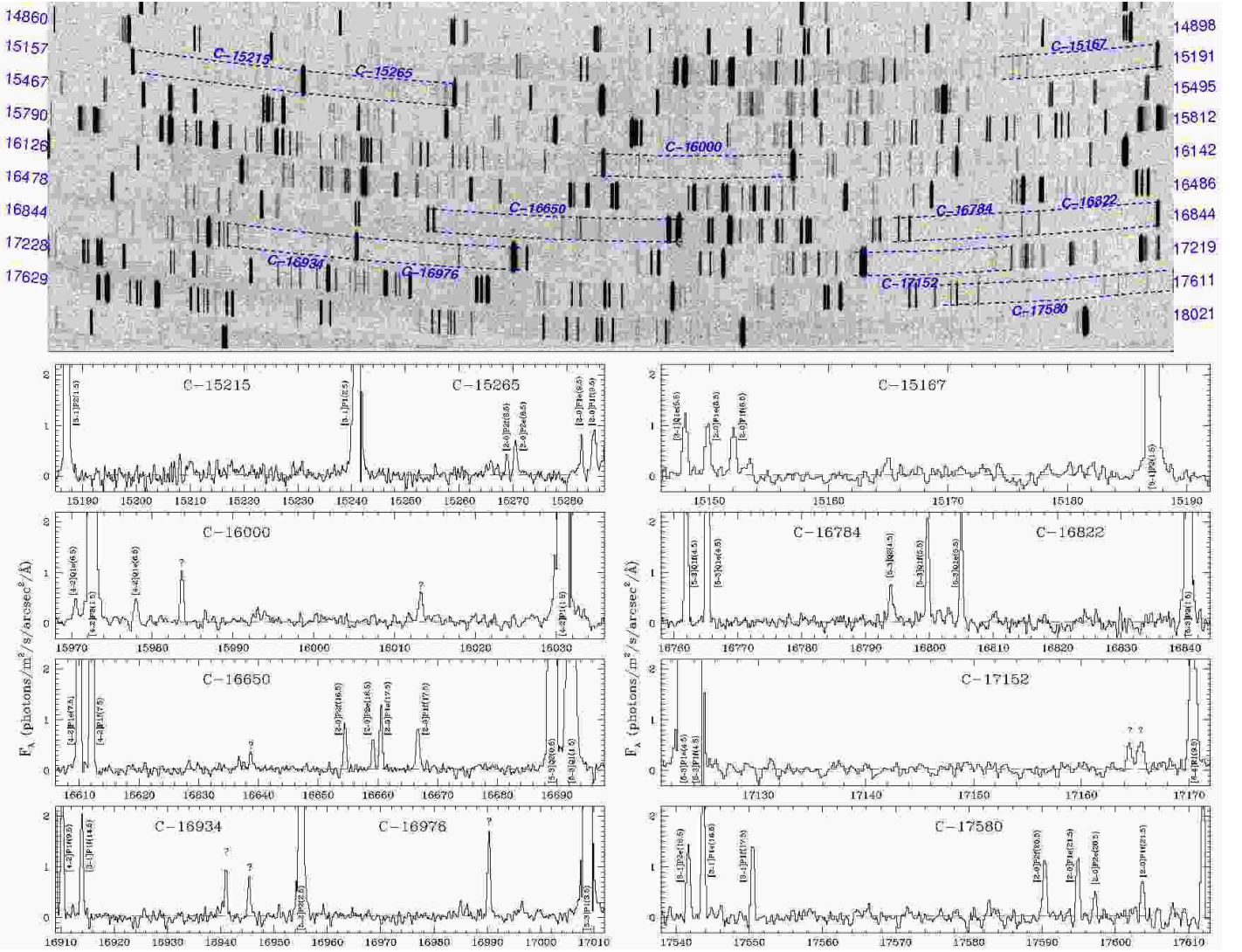}
      \caption{
           Upper panel: GIANO echelle spectrum of the H atmospheric band.
           Lower panels: extracted spectra in regions relatively free of
           line emission. The horizontal dashed lines show the level of
           300 photons/m$^2$/s/arcsec$^2$/$\mu$m
           (equivalent to 20.1 AB-mag/arcsec$^2$).
              }
         \label{fig_nice_H}
   \end{figure*}

The remaining features are not identified. Of these 34 lines
are closely spaced doublets with equal intensities. A representative
example are the lines at $\lambda\lambda$17164.5, 17165.5~\AA\ visible in
the lower-right panel of Fig.~\ref{fig_nice_H}. Several of these features were
already detected in Paper~1.
They are very similar to other $\Lambda$-split OH doublets
detected in our spectra. However, their wavelengths do not correspond
to any OH transition with $J_u\le40.5$ and $v_u\le10$.
The possibility that these doublets are produced by OH isotopologues
(e.g. $^{18}$OH) should be investigated, but is beyond the aims of
this paper.

   \begin{figure*}
   \includegraphics[width=\hsize]{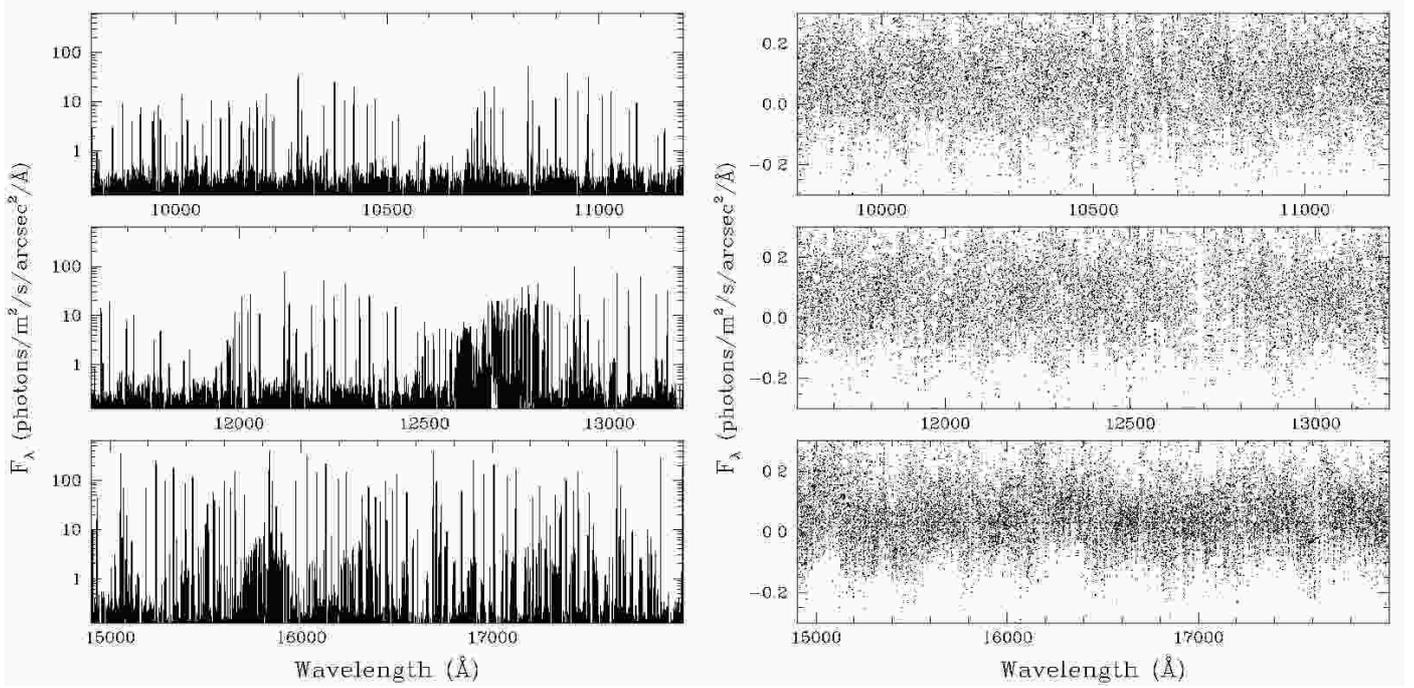}
      \caption{
        Left-hand panel: overview of the GIANO spectrum of the airglow.
        Right-hand panel: downscale to low flux levels. The spectral
        measurements are shown as separate dots to avoid confusion.
        The horizontal dashed line in the lowest panel shows the level of
           300 photons/m$^2$/s/arcsec$^2$/$\mu$m
           (equivalent to 20.1 AB-mag/arcsec$^2$).
        }
        \label{fig_show_all_and_zoom}
   \end{figure*}
\setcounter{table}{2}
\begin{table}
 \caption{Spectral bands with low contamination from lines}
 \label{table_selected_bands}
 \centering
 \begin{tabular}{|l|c|c|c|}
 \hline\hline
  Band & $\lambda$-range (\AA) &
    $\Delta\lambda/\lambda$ &
    Lines Flux\tablefootmark{(1)} \\
 \hline
  C-15167   & 15153 -- 15183 &  0.0020 & 0.6 (210\ ;\ 20.5) \\
  C-15215   & 15195 -- 15235 &  0.0026 & --        \\
  C-15265   & 15245 -- 15285 &  0.0026 & 1.6 (400\ ;\ 19.8) \\
  C-16000   & 15980 -- 16020 &  0.0025 & 1.2 (320\ ;\ 20.0) \\
  C-16650\tablefootmark{(2)}
            & 16620 -- 16680 &  0.0036 & 2.3 (380\ ;\ 19.9) \\
  C-16784   & 16770 -- 16798 &  0.0017 & 0.5 (170\ ;\ 20.7) \\
  C-16822   & 16808 -- 16836 &  0.0017 & --        \\
  C-16934   & 16918 -- 16950 &  0.0019 & 1.3 (390\ ;\ 19.8) \\
  C-16976   & 16962 -- 16990 &  0.0016 & 1.0 (360\ ;\ 19.9) \\
  C-17152   & 17134 -- 17170 &  0.0021 & 0.7 (190\ ;\ 20.6) \\
  C-17580   & 17555 -- 17605 &  0.0028 & 2.2 (430\ ;\ 19.7) \\
 \hline\hline
 \end{tabular}
 \tablefoot{ \ \\
  \tablefoottext{1}{
  First entry is the lines flux in photons/m$^2$/s/arcsec$^2$. Numbers in
   brackets are the equivalent continuum flux (i.e. the line flux averaged
   over the band-width) in photons/m$^2$/s/arcsec$^2$/$\mu$m
   and in AB-mag/arcsec$^2$.
   }\\
  \tablefoottext{2}{Region used by  Maihara et al. \cite{Maihara93}
   to define the sky-continuum}
  }
\end{table}

\subsection{The sky continuum emission}
\label{results_cont}

Within the H-band (1.5--1.8 $\mu$m) we detected:
\begin{itemize}
\item 514 lines of OH, half of which are produced by the 
hot-OH component described in Sect.~\ref{OH_hot};
\item 41 lines of O$_2$, including two broad and prominent band-heads;
\item 79 unidentified features.
\end{itemize}
Finding spectral regions free of emission features and far from bright airglow
lines is already difficult in our spectra. It becomes
virtually impossible at the lower resolving powers foreseen for MOONS
(R$\simeq$6,600) and other faint-object IR spectrometers.
In Fig.~\ref{fig_nice_H}
we show the observed 2D echellogram of GIANO and the extracted 1D spectra of
selected regions with relatively low contamination from lines. Their
main parameters are
listed in Table~\ref{table_selected_bands}. They were selected
with the following criteria:
\begin{itemize}
\item The width of the band must correspond to at least 10 resolution 
  elements of MOONS (i.e. $\Delta\lambda/\lambda\!>\!1/660$)
\item The band  must include only faint lines whose total flux,
  averaged over the band-width, is less than 500 
  photons/m$^2$/s/arcsec$^2$/$\mu$m
  (equivalent to 19.6 AB-mag/arcsec$^2$).
\end{itemize}
The broadest band
is C-16650. It coincides with the region used by Maihara et al.
\cite{Maihara93} to measure an average sky-continuum of
590 photons/m$^2$/s/arcsec$^2$/$\mu$m 
  (equivalent to 19.4 AB-mag/arcsec$^2$).
We find that about 65\% of this
flux can be ascribed to five emission features (4 lines from hot-OH
and one unidentified, see Fig.~\ref{fig_nice_H}) that lie close to
the centre of this band. 
Taken at face value, this would imply that the
true continuum is $\simeq$200 photons/m$^2$/s/arcsec$^2$/$\mu$m
  (equivalent to 20.6 AB-mag/arcsec$^2$).
However, this number is affected by large uncertainties intrinsic to the
procedure used to extract/average the continuum level from the spectrum and
to variations of the sky lines between different epochs. Indeed, to
reach a more reliable conclusion one would
should re-analyse the raw data of Maihara et al.~\cite{Maihara93} 
and correct them for the contribution of the sky-emission lines before
computing the continuum level.

We attempted to measure the sky continuum emission using
the extracted GIANO spectrum. This spectrum is shown in 
Fig.~\ref{fig_show_all_and_zoom} and listed in 
Table~4 (available only in electronic format). 
The H-band has enough S/N ratio to show a faint continuum  
of about 300 photons/m$^2$/s/arcsec$^2$/$\mu$m (equivalent to 20.1
AB-mag/arcsec$^2$); this level is shown as
a dashed line in the figure. 
It corresponds to 5 e$^-$/pixel/hr at the GIANO detector. A formal computation
of noise (i.e. including read-out, dark-current
and photon statistics) yields a convincing 5$\sigma$ detection
once the spectrum is re-sampled to a resolving power of R=5,000.
The contribution by systematic errors is more difficult to estimate. 
On the one hand, the procedure used to subtract detector dark and persistency 
(see Sect.~\ref{observations}) has
correctly produced a zero continuum in the bands where the
atmosphere is opaque (the uppermost order in the 2D frame of 
Fig.~\ref{fig_nice_H}).  
On the other hand, however, we cannot
exclude that second order effects have left some residual instrumental
artifacts in the H-band. 
An analysis of the dark frames affected by persistency indicates that
second order effects tend to increase the residuals, rather than 
over-subtracting the residual continuum level in the H-band.
Therefore, we are reasonably confident that the true sky-continuum
cannot be larger than the observed value. 

In the Y and J bands our spectra have a lower S/N ratio
because the efficiency of the GIANO detector drops at shorter wavelengths. 
The measured upper limits correspond to about 19 AB-mag/arcsec$^2$ and
are compatible with the measurements by Sullivan \& Simcoe \cite{Sullivan}.
In general, the Y and J bands are much less contaminated by line emission 
and the higher resolving power of GIANO is no longer
needed to find spectral regions that properly sample the sky continuum.

\section{Discussion and conclusions}
\label{discussion}

We took advantage of the second commissioning of the GIANO high-resolution
infrared spectrograph at La Palma Observatory
to point the instrument directly to the sky.
This yielded a sky spectrum much deeper than those collected through the
fibre-interface to the TNG telescope and published in Oliva et al.~\cite{paper1}.
The spectrum extends from 0.97 to 2.4 $\mu$m and includes the whole
Y, J, and H-bands. 

The spectrum shows about 1500 emission lines, a factor of two more than in
previous works. Of these, 80\% are identified as
OH transitions while the others are attributable to O$_2$ or unidentified.
Roughly half of the OH lines arise from highly excited rotational
states, presumably associated with lower density clouds at higher altitudes.
We derive physical parameters useful to model this hot-OH component that
as yet has never been included in the airglow models used by astronomers.

Several of the faint lines are in spectral regions that
were previously believed to be free of lines emission.
The continuum in the H-band is marginally detected
at a level of about 300 photons/m$^2$/s/arcsec$^2$/$\mu$m
equivalent to 20.1 AB-mag/arcsec$^2$.
In spite of the very low sky-continuum level, the myriad of airglow 
emission lines in the H-band severely limits 
the spectral ranges that can be
properly exploited for deep observations of faint objects
with low/medium resolution spectrographs.
We have identified a few spectral bands that could still remain quite dark
at the resolving power foreseen for the faint-object spectrograph
VLT-MOONS (R=6,600).

The spectrum and the
updated lists of observed infrared sky-lines are published in electronic
format.

\begin{acknowledgements}
  Part of this work was supported by the grants ``TECNO-INAF-2011''
  and ``Premiale-INAF-2012".
\end{acknowledgements}


\begin{thebibliography}{}

  \bibitem[2009]{Bernath}  
  Bernath, P.\ F., \& Colin, R.\  2009, J. Mol. Spec., 257, 20

  \bibitem[2004]{blandhawthorn04}  
   Bland-Hawthorn, J., Englund, M., \& Edvell, G.\ 2004,
    Optics Express, 12, 5902

  \bibitem[2011]{cirasuolo11}  
  Cirasuolo, M., Afonso, J., Bender, et al.\ 2011, The Messenger, 145, 11

  \bibitem[2014]{cirasuolo14}  
  Cirasuolo, M., Afonso, J., Carollo, M., et al. \ 2014, SPIE, 9147E, 0N1

  \bibitem[1996]{Content96} 
   Content, R.\ 1996, \apj, 464, 412

  \bibitem[2007]{Cosby} 
    Cosby, P.\ C., \& Slanger, T.\ G.\ 2007, Can. J. Phys., 85, 77

  \bibitem[2000]{cuby00} 
    Cuby, J.\ G., Lidman, C., \& Moutou, C.\  2000, Messenger, 101, 2

  \bibitem[2012]{ellis12} 
   Ellis, S.\ C., Bland-Hawthorn, J., Lawrence, J.,
   et al.\ 2012, \mnras, 425, 1682

  \bibitem[1998]{Ennico98}  
   Ennico, K.\ A., Parry, I.\ R., Kenworthy, M.\ A., et al.\ 1998,
   SPIE, 3354, 668

  \bibitem[1994]{Herbst94}  
    Herbst, T. M. \ 1994, \pasp, 106, 1298

  \bibitem[2001]{Iwamuro01}  
   Iwamuro, F., Motohara, K., Maihara, T., Hata, R.,
   \& Harashima, T.\ 2001, \pasj, 53, 355

  \bibitem[2006]{Iwamuro06}  
   Iwamuro, F., Maihara, T., Ohta, K., et al.\  2006, SPIE, 6269, 1B1

  \bibitem[2014]{licausi}  
  Li Causi, G., Cabral, A., Ferruzzi, D., et al. \ 2014 SPIE, 9147E, 641

  \bibitem[1993]{Maihara93} 
   Maihara, T., Iwamuro, F., Yamashita, T., et al.\ 1993, \pasp, 105, 940

  \bibitem[1992]{Oliva92} 
  Oliva, E., \& Origlia, L.\ 1992, \aap, 254, 466

  \bibitem[2012a]{oliva_spie2012_1}
  Oliva, E., Origlia, L., Maiolino, R., et al.\ 2012a, SPIE, 8446, 3T1

  \bibitem[2012b]{oliva_spie2012_2}
  Oliva, E., Biliotti, V., Baffa, C.,  et al.\ 2012b, SPIE, 8453, 2T1

  \bibitem[2013]{paper1}
  Oliva, E., Origlia, L., Maiolino, R., et al.\ 2013, \aap, 555, A78

  \bibitem[2011]{redman2011} 
    Redman, S.\ L., Lawler, J.\ E., Nave, G., Ramsey, L.\ W., \& Mahadevan, S.\
    2011, \apjs, 195, 24

  \bibitem[2009]{hitran} Rothman, L.\ S., Gordon, I.\ E., Barbe, A., et al.\
        2009, JQS\&RT, 110, 533

  \bibitem[2000]{Rousselot00} 
    Rousselot, P., Lidman, C., Cuby, J.-G., Moreels , G., \& Monnet, G.\
    2000, \aap, 354, 1134

  \bibitem[1985]{Sharma} 
    Sharma, R.\ D.\  1985, Handbook of Geophysics, chapter 13 (Air Force
        Geophysics Laboratory, USAF)

  \bibitem[2012]{Sullivan} 
    Sullivan, P.\ W., \& Simcoe, R.\ A.\ 2012, \pasp, 124, 1336

  \bibitem[2014]{tozzi14} 
    Tozzi, A., Oliva, E., Origlia, L., et al.\ 2014, SPIE, 9147E, 9N1

  \bibitem[2013]{Trinh13}  
   Trinh, C.\ Q., Ellis, S.\ C., Bland-Hawthorn, J., et al.\
   2013, \mnras, 432, 3262

  \bibitem[2007]{vanderloo} 
    van der Loo, M.\ P.\ J., \& Groenenboom, G.\ C.\ 2007,
     J. of Chemical Physics, 126, 114314
%
\end{thebibliography}
\end{document}